\documentclass[reprint, amsmath,amssymb, aps]{revtex4-2}

\usepackage{graphicx}  
\usepackage{bm}        
\usepackage{hyperref}
\usepackage{xcolor}

\newcommand{\aver}[1]{ \! \left\langle {#1} \right \rangle \!}
\newcommand{\vect}[1]{\boldsymbol{#1}}

\begin{document}



\title{Anisotropic mean flow enhancement and anomalous transport \\ of finite-size spherical particles in turbulent flows}
\author{Alessandro Chiarini}
\author{Ianto Cannon}
\author{Marco Edoardo Rosti}
\email[]{marco.rosti@oist.jp}
\affiliation{Complex Fluids and Flows Unit, Okinawa Institute of Science and Technology Graduate University, 1919-1 Tancha, Onna-son, Okinawa 904-0495, Japan.}   
                          
\date{\today}

\begin{abstract}

We investigate the influence of dispersed solid spherical particles on the largest scales of the turbulent Arnold-Beltrami-Childress (ABC) flow. The ABC flow is an ideal instance of a complex flow: it does not have solid boundaries, but possesses an inhomogeneous and three-dimensional mean shear. By tuning the parameters of the suspension, we show that particles modulate the largest scales of the flow towards an anisotropic, quasi-two-dimensional and more energetic state. In this regime, particles move along quasi-straight trajectories and exhibit anomalous transport.

\end{abstract}

\maketitle

Particle-laden turbulent flows have attracted the attention of many scholars over the last decades. Their significance goes beyond a fundamental interest and encompasses several applications 
 such as blood flow in the human body, the food industry, and pyroclastic flows \citep{delillo-etal-2014,breard-etal-2016,falkinhoff-etal-2020}. Also, the modulation of the mean flow and the enhancement/attenuation of turbulence due to particles are relevant in both environmental \citep{sengupta-etal-2017} and industrial flows \citep{ferreyra-etal-2011}. However, although this is a classical problem in fluid mechanics since the seminal work by \citet{tsuji-morikawa-1982}, the multi-scale mechanism governing the fluid-particle interaction is still an open question. In particular, the ability of suspensions of solid particles to modify and control the largest scales of a generic and complex flow is unclear.

The presence of the solid phase alters the momentum of the flow, and may result in modulation of the 
carrier fluid \citep{balachandar-eaton-2010,brandt-coletti-2022}. When the suspension is dilute enough, the fluid phase can be considered unaltered by the presence of the particles. Instead, when the suspension is non-dilute, the fluid phase undergoes macroscopic changes in a way that depends on several parameters, such as, for example, the size and density of the particles, and the volume and mass fractions of the suspension \citep{gore-crowe-1989,elgobashi-2006}. 
Over the last years, the solid-fluid interaction in particle-laden turbulent flows has been the subject of several studies in various flows, ranging from homogeneous and isotropic flows \citep{oka-goto-2022} to wall-bounded flows \citep{costa-etal-2016}.
 Nevertheless, the influence of the solid phase on the largest scales of a generic and complex flow has not yet been satisfactorily addressed, and the accurate characterisation of the underlying physics still requires significant effort. Here, we aim to address the following questions: \textit{How do suspensions of solid particles modulate the largest and most energetic scales of the flow in the presence of an inhomogeneous mean shear? Is it possible to use particles to effectively modify and control the mean flow?}

\begin{figure}
\centering
\includegraphics[width=0.45\textwidth]{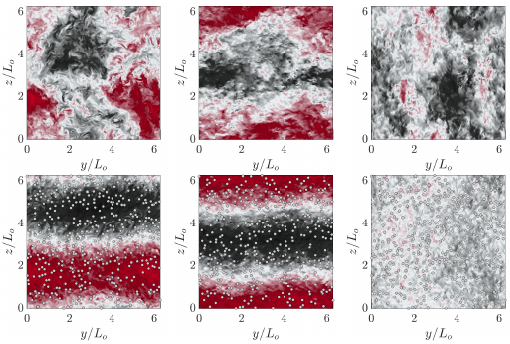}
\caption{Instantaneous velocity field in the $x=L/2$ plane for the (top) single-phase case and the (bottom) particulate case with $D/L \approx 0.02$ and $M=0.6$. Left: $u$; centre: $v$; right: $w$. The symmetric black-to-red colourmap goes from $-9 \le u_i/U_o \le 9 $ \cite{Note2}. We denote with $z$ the direction orthogonal to the trajectories' plane \cite{Note3}.}
\label{fig:inst}
\end{figure}
In order to answer these questions, in this letter we study the modulation of the turbulent Arnold-Beltrami-Childress (ABC) flow by finite-size particles, with a focus on the largest scales. In a Cartesian reference system the laminar ABC flow is $L-$periodic in the three $x$, $y$ and $z$ directions, with a velocity field $\vect{u} = (u,v,w)$ that depends on four real parameters $V_o$, $A$, $B$ and $C$, i.e.
\begin{equation}
\begin{cases}
u/V_o &= A \sin \left( \frac{2 \pi}{L} z \right) + C \cos \left( \frac{2 \pi}{L} y \right), \\
v/V_o &= B \sin \left( \frac{2 \pi}{L} x \right) + A \cos \left( \frac{2 \pi}{L} z \right), \\
w/V_o &= C \sin \left( \frac{2 \pi}{L} y \right) + B \cos \left( \frac{2 \pi}{L} x \right);
\end{cases}
\label{eq:ABC}
\end{equation}
$A$, $B$, and $C$ are bounded between $0$ and $1$ and determine the shape of the flow, while $V_o$ determines the magnitude. Here, we consider the turbulent ABC flow, which is obtained by forcing the incompressible Navier--Stokes equations with the ABC forcing with $A=B=C=1$ \cite{podvigina-pouquet-1994} (see figure \ref{fig:inst}). This provides an ideal instance of a complex turbulent flow with a three-dimensional (3D) mean field and an inhomogeneous mean shear, such as flow past an object or in a curved pipe. Because of the lack of solid boundaries, in the ABC flow the complex flow structures induced by the presence of a wall are avoided. Thus the particle-laden turbulent ABC flow allows us to isolate the influence of the particles on the large-scale motions, that in wall-bounded flows might be hidden by the complex near-wall phenomenology. In this idealised framework, we show that 
 non-dilute suspensions of solid particles can substantially modify the structure of the mean flow. When tuning their size and density, indeed, particles 
 modulate the largest scales of the flow towards an anisotropic, almost two-dimensional (2D) and more energetic state (see figure \ref{fig:inst}). Intriguingly, we show that this happens in the presence of a sustained 3D external forcing, whose effect is overcome by the presence of the solid phase. This paves the way for the use of solid particles to control complex 3D flows.

To tackle this problem, we have performed 3D direct numerical simulations of the flow within a triperiodic box of size $L$, with dispersed particles of various finite sizes, that lie within the inertial range of turbulence. The fluid and the solid dynamics are fully resolved, and coupled with an immersed boundary method \citep{hori-rosti-takagi-2022}. The external ABC forcing is set to achieve, in the single-phase case, a micro-scale Reynolds number of $Re_\lambda = u' \lambda/ \nu \approx 435$, where $u'$ is the root mean square of the velocity fluctuations and $\lambda$ is the Taylor length scale~\footnote{See the Supplemental Material for more information, with Refs.\cite{eswaran-pope-1988,tsuji-etal-1993,pope-2000,kajishima-etal-2001,monti-etal-2021}.}. 
The particle diameter is varied between $0.0104 \le D/L \le 0.0796$.
For each particle size, the number of particles is set to provide a volume fraction of $\Phi_V \approx 0.08$, which is large enough for the suspension to be non-dilute and small enough for the particle-particle interactions to be sub-dominant. Finally, the ratio between the density of the particles and the fluid is varied between $1.3 \le \rho_p/\rho_f \le 105$, to consider both light and heavy particles, yielding a variation of the mass fraction between $0.1 \le M \le 0.9$. 

\begin{figure}
\centering
\includegraphics[width=0.45\textwidth]{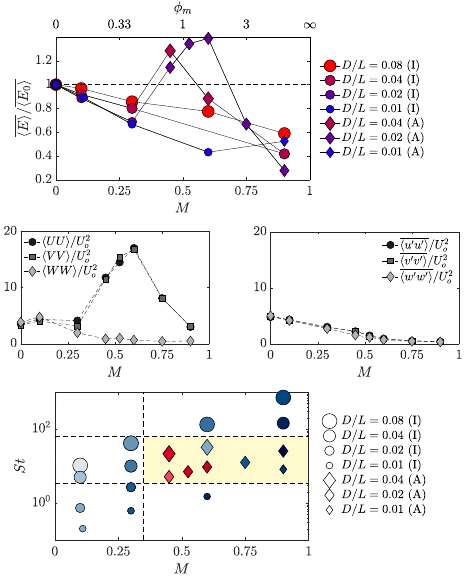}
\caption{Modulation of the flow energy by finite-size solid particles. Top panel: total kinetic energy of the flow as a function of the mass fraction $M$ and of the mass loading $\phi_m$, for different particle size. $E_0$ is the total kinetic energy of the single-phase flow. Circles and diamonds refer to the isotropic (I) and anisotropic (A) regimes. In the central panels the influence of the solid phase on the mean and fluctuating field is isolated for $D/L\approx 0.02$. Centre left: variance of the mean velocity components. Centre right: variance of the fluctuating velocity components. Bottom panel: modulation of the kinetic energy of the flow as a function of $M$ and Stokes number $St$. Blue symbols are for $\overline{\aver{E}}<\overline{\aver{E_0}}$, red symbols for $\overline{\aver{E}}>\overline{\aver{E_0}}$ with the divergent blue-to-red colourmap being for $0.25 \le \overline{\aver{E}}/\overline{\aver{E_0}} \le 1.75$. The shaded region shows the anisotropic regime.}
\label{fig:TotEner}
\end{figure}
Figure \ref{fig:TotEner} shows that the modulation of the carrier ABC flow changes with the size and density of the particles. In the top panel, we plot the total kinetic energy of the flow $\overline{ \aver{E(\vect{x},t)} }$ for different values of $D$ and $M$, where $E(\vect{x},t) = | \vect{u}(\vect{x},t) |^2/2$, while $\aver{\cdot}$ and $\overline{\cdot}$ indicate average along homogeneous directions and in time, respectively. Overall, the total energy
decreases while reducing $D$ and increasing $M$, in agreement with the findings of previous authors \citep{tenCate-etal-2004,uhlmann-chouippe-2017,olivieri-cannon-rosti-2022,oka-goto-2022}. However, the trend is non-monotonic, and the large increase of $\overline{\aver{E}}$ for $M \gtrsim 0.45$ and $D/L \lesssim 0.04$ shows that the solid phase modulates the carrier flow in a way that sharply changes with the density of the particles. Interestingly, solid suspensions of particles with $0.02 \lesssim D/L \lesssim 0.04$ and $0.45 \le M \le 0.6$ enhance the total energy of the carrier flow with respect to the single-phase case. In the central panels of figure \ref{fig:TotEner}, we use a temporal average to isolate the influence of the solid phase on the large and small scales of the carrier flow. We consider the $D/L \approx 0.02$ particulate cases and, after decomposing the velocity field into its temporal mean $ \vect{U}(\vect{x}) = \overline{\vect{u}(\vect{x},t)}$ and fluctuations $ \vect{u}'(\vect{x},t) = \vect{u}(\vect{x},t) - \vect{U}(\vect{x})$, we plot the variances of their three components~\footnote{All quantities are made dimensionless with $L_o$ and $U_o$, where $L_o = L/2\pi$ and $U_o=\sqrt{F_o L_o}$ with $F_o$ denoting the intensity of the ABC forcing; see the Supplemental Material \cite{Note1}.}. Light particles ($M \le 0.3$) modulate the flow without introducing a preferential direction, and attenuate the energy of the fluid phase by modifying the mean and fluctuating fields in an isotropic way. Indeed, here $\aver{UU} \approx \aver{VV} \approx \aver{WW}$ and $\overline{\aver{u'u'}} \approx \overline{\aver{v'v'}} \approx \overline{\aver{w'w'}}$. Heavier particles ($M \ge 0.45$), instead, modulate differently the mean and fluctuating fields. They continue to attenuate the energy of the fluctuations in an isotropic way, but modulate the mean field towards an anisotropic, quasi-2D and more energetic state. In this case, $\aver{UU} \approx \aver{VV} \gg \aver{WW}$~\footnote{Note that we denote with $z$ the direction aligned with the mean-flow component that is attenuated by the particle modulation, corresponding to the direction orthogonal to the particle trajectories' plane. See the Supplemental Material \cite{Note1}.}, while $\overline{\aver{u'u'}} \approx \overline{\aver{v'v'}} \approx \overline{\aver{w'w'}}$. The increase of the total flow energy observed for $0.45 \le M \le 0.6$ is thus entirely due to the mean-flow energy enhancement. 

The anisotropic state is observed in a limited portion of the space of mass fraction $M$ and Stokes number $St$, 
i.e. $5 \lesssim St \lesssim 60$ and $M \gtrsim 0.4$ (see the bottom panel of figure \ref{fig:TotEner}). 
The anisotropy of the large scales is indeed favoured by the motion of the particles when, due to their inertia, they tend to move along straight trajectories. 
Particles with $St \lesssim 5$ follow the complex fluid motion and hence do not significantly modify the flow. However, for $5 \lesssim St \lesssim 60$, inertia causes the particle trajectories to straighten, causing the fluid motion to reorganise into a quasi-2D state. When $St$ further increases, the particles become less responsive to the fluid motion, and the 2D state does not develop. 
Note that the isotropic flow modulation observed for $St \gtrsim 60$ is consistent with the limit of infinite inertia: when $St \rightarrow \infty$, the particles do not move, and the quasi-2D flow does not arise. The dependence of the flow modulation on $M$ shows that the emergence of the anisotropic state requires a strong enough backreaction of the particles to the fluid phase. Also, the ability of the particles to modulate the largest scales of the carrier flow strongly depends on the ratio $D/L$. In fact, the strongest flow modulation (and largest mean-flow enhancement) is observed for the intermediate $D/L\approx0.02$, while the effect is lower for both smaller and larger particles. For $D \ll L$, indeed, particles have a small inertia, follow the fluid and do not favour the flow two-dimensionalisation. For $D\approx L$, instead, the motion of the particles is only marginally influenced by the mean shear, as it is mainly driven by fluid velocity fluctuations with a length scale larger than their characteristic size. See the Supplemental Material \cite{Note1} for further discussions, and for the dependence of the anisotropic flow modulation on the inhomogeneous mean shear.

\begin{figure}
\centering
\includegraphics[width=0.45\textwidth]{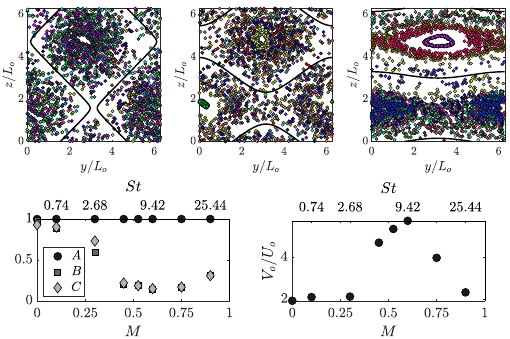}
\caption{Mean flow field for $D/L\approx 0.02$. Top: Poincar\'{e} section for the mean flow of the single-phase case (right) and of the particulate cases with $M=0.3$ (centre) and $M=0.6$ (right). Twenty streamlines are considered, each identified by a different colour. 
 Bottom: variation with $M$ and $St$ of $A$, $B$, $C$ and $V_o$ used in equation \ref{eq:ABC} to approximate the mean flow. Without losing generality we set that the maximum between $A$, $B$ and $C$ is equal to $1$ \cite{Note1}. The black lines in the Poincar\'{e} sections are for $U=A \sin(z) + C \cos(y)=0$ with $A$ and $C$ as in the bottom left panel.}
\label{fig:Pmap}
\end{figure}
%
Next, we use Poincar\'{e} sections \cite{poincare-1982} to characterise the modulation of the mean flow $\bm{U}$.  
These are a 2D coding of a 3D dynamical system, in which one represents only the successive intersections of the mean-flow streamlines with one plane of the triperiodic domain. Figure \ref{fig:Pmap} shows the $x=L/2$ Poincar\'{e} section for the 
mean flow of the turbulent single-phase case, and of the $D/L\approx0.02$ particulate cases with $M=0.3$ and $M=0.6$. For each streamline many successive intersections with the $x=L/2$ plane are represented. 
At first glance, the Poincar\'{e} maps show that points associated with the same streamline are found to span a narrower range of $z$ as the flow anisotropy increases. This confirms that in the anisotropic regime, particles attenuate the $W$ velocity component and modulate the mean field towards a quasi 2D state (see figure \ref{fig:TotEner}). The density of the points in the Poincar\'{e} section is proportional to the magnitude of the mean velocity component aligned with the direction perpendicular to the plane \citep{dombre-etal-1986}. In the laminar case, indeed, the density of the points in the $x=L/2$ Poincar\'{e} plane falls to zero along the $U/V_o= \sin(z/L_o) + \cos(y/L_o) = 0$ line (see figure 5 of the Supplemental Material \cite{Note1}). A similar pattern is observed for the turbulent single-phase case. In fact, a remarkable property of the ABC flow is that even in the turbulent regime, the mean velocity has nearly the laminar ABC profile, as observed by others for the Kolmogorov flow \cite{borue-orszag-1996,musacchio-boffetta-2014}. In the particulate cases, the scenario progressively changes with $M$. Interestingly, the solid phase modulates the mean flow in a way that it resembles a generic ABC flow, but with different values of $A$, $B$, $C$ and $V_o$. In view of this, the bottom panels of figure~\ref{fig:Pmap} report the coefficients of the generic ABC flow that best approximates the mean flow for $D/L\approx 0.02$ and $0.1 \le M \le 0.9$ \cite{Note1}. As $M$ (and $St$) increases, $A=1$ remains constant, while $B \approx C$ decrease and reach a minimum for $M=0.6$ ($St \approx 9.42$) where the flow anisotropy is maximum. In this case, $A \gg B \approx C$ and the streamlines of the mean flow are almost straight lines that lay in the $x-y$ planes, as $(U,V,W) \sim \left( \sin(z/L_o), \cos(z/L_o), 0 \right)$. Thus, particles with $0.45 \le M \le 0.9$ and $5.13 \lesssim St \lesssim 25.45$ attenuate the $W$ velocity component, and the mean flow approaches an anisotropic and 2D state, almost losing its dependence on $x$ and $y$; see the black lines in figure \ref{fig:Pmap}. $V_o$ sharply increases for $M \ge 0.45$ ($St \gtrsim 5.13$), in agreement with the enhancement of the mean-flow energy shown in figure \ref{fig:TotEner}. This anisotropic and almost 2D state resembles the bifurcated $\mathcal{A}_2$ state of the single-phase laminar ABC flow found by Podvigina and Pouquet \cite{podvigina-pouquet-1994}, spontaneously emerging at lower Reynolds numbers. Thus, it appears that the presence of particles with $5 \lesssim St \lesssim 60$ and $M > 0.3$, and their tendency to follow straight trajectories, changes the stability of the system, enabling thus the occurrence of an anisotropic $\mathcal{A}_2$-like state at the large scales also for these larger $Re$ \cite{Note1}. 

\begin{figure}
\centering
\includegraphics[width=0.45\textwidth]{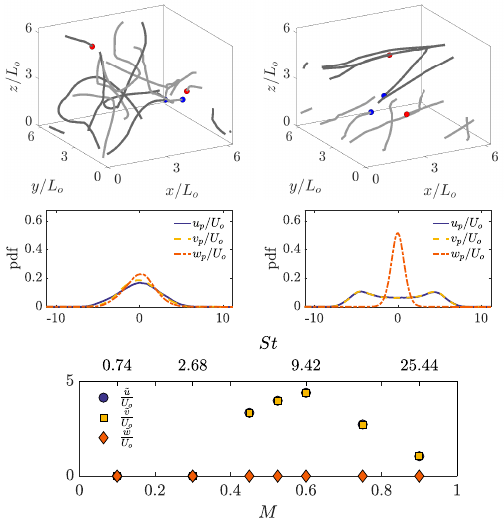}
\caption{Particle dynamics for $D/L\approx0.02$. Top panels: two representative trajectories for (left) $M=0.3$ and (right) $M=0.6$; the red/blue circle indicate the beginning/end of the trajectories. Central panels: probability density functions for (left) $M=0.3$ and (right) $M=0.6$. Bottom panel: dependence of the modes $\tilde{u}_i$ of the probability density functions on $M$ and $St$.}
\label{fig:VelPar}
\end{figure}

\begin{figure}
\centering
\includegraphics[width=0.45\textwidth]{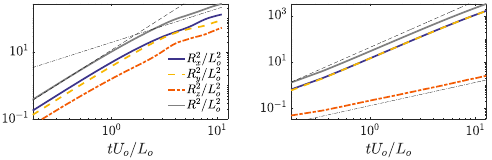}
\caption{Particle mean squared displacement $R^2(t)$ for $D/L\approx 0.02$. Left: $M=0.3$. Right: $M=0.6$. The thin dashed line represents $t^2$, while the thin dash-dotted line $t$.}
\label{fig:mds}
\end{figure}

To further investigate the fluid-solid interaction, figures \ref{fig:VelPar} and \ref{fig:mds} characterise the particles' dynamics for $D/L\approx0.02$. In the isotropic regime ($M \le 0.3$ and $St \lesssim 2.68$), particles are able to follow the intricate ABC cellular forcing, and exhibit complex trajectories as expected by the chaotic Lagrangian structure of the ABC flow \cite{dombre-etal-1986}. In this case, the particles do not select a preferential direction, and explore the entire computational domain. Indeed, the probability density functions (pdfs) of the three velocity components almost collapse, exhibiting a symmetric unimodal distribution. 
 On the contrary, in the anisotropic regime ($ 0.45 \le M \le 0.9$ and $5.13 \lesssim St \lesssim 25.44$), the particles move along almost straight trajectories with $u_p \approx v_p \gg w_p$. 
 The direction of the trajectories of the particles in the $x-y$ plane changes with $z$, in agreement with the fact that $A \gg B \approx C$, and $U \sim \sin(z/L_o)$ and $V \sim \cos(z/L_o)$. As a result, the in-plane components of the particle velocity exhibit a symmetric bimodal distribution, and the modes $\pm \tilde{u}_p$ and $\pm \tilde{v}_p$ change with the density of the particles (see bottom panel of figure \ref{fig:VelPar}).

Figure \ref{fig:mds} shows the mean squared displacement of the particles, $R(t)^2 = \aver{|\bm{x}_p(t) - \bm{x}_p(0)|^2}_p$
; here, $\aver{\cdot}_p$ indicates average over particles. 
The particle motion is diffusive (ballistic) when $R^2(t) \sim t^\alpha$ with $\alpha=1$ ($2$). 
 As expected, in the isotropic regime, particles show ballistic motion for short times, while at large times, their motion becomes uncorrelated under the action of the random velocity fluctuations, and they exhibit a diffusive process; see the $M=0.3$ case in the left panel. In the anisotropic regime, instead, particles exhibit anomalous transport with $1 < \alpha < 2$ at large times; see the $M=0.6$ case in the right panel. The scaling of the anomalous transport slightly changes with the density of the particles in the $1.6 \lessapprox \alpha \lessapprox 1.75$ range, increasing as the particles' motion becomes more coherent and their trajectories more straight. Note that considering the $z$ direction alone, i.e. $R_z(t)^2 = \aver{|z_p(t)-z_p(0)|^2}_p$, the diffusive transport with $\alpha \approx 1$ is recovered also for heavy particles.
 
To conclude, we have shown that in the presence of a 3D and inhomogeneous mean shear, solid particles can substantially modify the structure of the largest scales of the flow. By tuning the size and the density of the particles, non-dilute suspensions can modulate the largest scales of the flow towards an anisotropic, almost 2D and more energetic state, while preserving isotropy at smaller scales. 
The ability of solid particles to modify the large-scale flow may pave the way for their use in flow-control purposes. For example, the cooling and cleaning effects of oil in an engine are undermined by recirculating regions in the flow~\citep{Gamble_Priest_Taylor_2003,concli_advanced_2023}. We showed that adding particles can potentially eliminate such recirculating regions, improving engine efficiency and lifetime. Also, our results show that there is a specific regime in which the ballistic motion of the particles is favoured over diffusive motion, with relevance for drug delivery and pollen dispersal~\citep{hamaoui-laguel_effects_2015,ito_ballistic_2023}.

\begin{acknowledgments}
The research was supported by the Okinawa Institute of Science and Technology Graduate University (OIST) with subsidy funding from the Cabinet Office, Government of Japan. The authors acknowledge the computer time provided by the Scientific Computing section of Research Support Division at OIST.
\end{acknowledgments}

%


\end{document}


\title{SUPPLEMENTAL MATERIAL\\[8pt]
Anisotropic mean flow enhancement and anomalous transport of finite-size spherical particles in turbulent flows}

\author{Alessandro Chiarini, Ianto Cannon and Marco Edoardo Rosti}

%

\maketitle

\onecolumngrid


\section{Governing equations}

This section provides information on the physical model considered in the present study. We first introduce the governing equations for the fluid flow, and then we provide an overview of how the solid phase is modelled. 

\subsection{The carrier flow}
 
We consider the turbulent 1:1:1 ABC flow, i.e., a Newtonian fluid that obeys the well-known incompressible Navier-Stokes equations
\begin{equation}
\begin{cases}
\frac{\partial \bm{u}}{\partial t} + \left( \bm{u} \cdot \bm{\nabla} \right) \bm{u} = - \frac{1}{\rho_f} \bm{\nabla} p + \nu \bm{\nabla}^2 \bm{u} + \bm{f} +  \bm{f}^{\leftarrow p},\\
\bm{\nabla} \cdot \bm{u} = 0,
\end{cases}
\end{equation}
where $\bm{u}=\bm{u} \left( \bm{x},t \right)$ and $p=p\left( \bm{x},t \right)$ are the velocity and pressure fields, $\rho_f$ and $\nu$ are the fluid volumetric density and kinematic viscosity, and $\bm{f}^{\leftarrow p}$ is a singular force distribution used to mimic the presence of the dispersed solid objects by indirectly imposing the no-penetration and no-slip conditions at the fluid-solid interface. The Navier--Stokes equations are forced with the ABC forcing \cite{podvigina-pouquet-1994}, i.e.,
\begin{equation}
  \bm{f} = F_o 
  \begin{pmatrix}
  A \sin \left( 2 \pi z/L \right) + C \cos \left( 2 \pi y/L \right) \\
  B \sin \left( 2 \pi x/L \right) + A \cos \left( 2 \pi z/L \right) \\
  C \sin \left( 2 \pi y/L \right) + B \cos \left( 2 \pi x/L \right) \\
  \end{pmatrix}
\end{equation}
where $A=B=C=1$, $L$ is the forcing wavelength, and $F_o$ is a constant that determines its intensity. The forcing is set to achieve a Reynolds number of $Re = U_o L_o/ \nu \approx 894$, where $L_o$ and $U_o$ are the reference length and velocity scales, defined as $L_o= L/2 \pi$ and $U_o = \sqrt{F_o L_o}$. In the single-phase configuration, this leads to a micro-scale Reynolds number of $Re_\lambda = u' \lambda / \nu \approx 435$, where $u'$ is the root-mean-square of the velocity fluctuations, and $\lambda$ is the Taylor length scale.  

\begin{figure}
\centering
\includegraphics[width=0.98\textwidth]{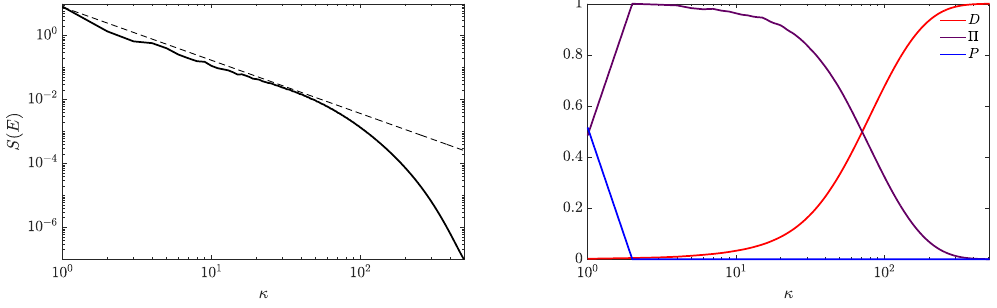}
\caption{Left: Energy spectrum for the single-phase configuration. The dashed line denotes the $-5/3$ power law predicted by the Kolmogorov theory. Right: energy-transfer balance for the single-phase configuration. The terms in the energy budget are normalised with the dissipation rate $\epsilon$.}
\label{fig:Spec}
\end{figure}

Figure \ref{fig:Spec} depicts the resulting energy spectrum (left) and energy-transfer balance for the single-phase configuration (right).
The energy-transfer balance derives after manipulation of the Navier--Stokes equations \citep{pope-2000} and reads:
\begin{equation}
P(\kappa) + \Pi(\kappa) + D(\kappa) = \epsilon,
\end{equation}
where
\begin{subequations}
\begin{align}
P(\kappa)  &=\int_{\kappa}^\infty \frac{1}{2} \left( \hat{\bm{f}} \cdot \hat{\bm{u}}^* + \hat{\bm{u}} \cdot \hat{\bm{f}}^* \right) \text{d} \tilde{\kappa},\\
\Pi(\kappa)&=\int_{\kappa}^\infty - \frac{1}{2} \left(\hat{\bm{G}} \cdot \hat{\bm{u}}^* + \hat{\bm{u}} \cdot \hat{\bm{G}}^* \right) \text{d} \tilde{\kappa},\\
D(\kappa) &= \int_0^\kappa \left( - 2 \nu \tilde{\kappa}^2 \hat{E} \right) \text{d} \tilde{\kappa}.
\end{align}
\end{subequations}
Here $\hat{\cdot}$ denotes the Fourier transform, the $*$ superscript denotes complex conjugate, $\hat{\bm{G}}$ is the Fourier transform of the non linear term $\bm{u} \cdot \bm{ \nabla} \bm{u}$, and $\epsilon$ is the dissipation. As clearly visible by means of the energy spectrum (left) and of the plateau of the non-linear flux $\Pi$ (right), the Reynolds number considered in this work leads to a proper scale separation, with an inertial range that extends to almost two decades of wavenumbers.

\subsection{The solid phase}

We consider suspensions of solid spherical particles. The velocity at an arbitrary point $\bm{x}$ inside the particle, $\bm{u}_p \left( \bm{x}(t), t \right)$, can be described using the translational velocity, $\bm{v}_p$, and the rotational velocity, $\bm{\omega}_p$, of its centre of mass $\bm{x}_p$, as
\begin{equation}
  \bm{u}_p = \bm{v}_p + \bm{\omega}_p \times \bm{r},
\end{equation}
where $\bm{r}= \bm{x}- \bm{x}_p$ is the vector going from $\bm{x}_p$ to $\bm{x}$. The time evolution of the spherical particles is, therefore, based on the classical Newton-Euler equations for rigid body dynamics that read
\begin{equation} \label{eq:newtoneuler}
\begin{cases}
  m_p \frac{\text{d} \bm{v}_p}{\text{d} t} &= \bm{f}^{\leftarrow f} + \bm{f}^{\leftrightarrow p}, \\
  I_p \frac{\text{d} \bm{\omega}_p}{\text{d} t} &= \bm{L}_p^{\leftarrow f},
\end{cases}
\end{equation}
where $m_p = \pi \rho_p D^3/6$ and $I_p = m_p D^2/10$ are the mass and inertial moment of the particle, with $\rho_p$ being the particle density and $D$ the particle diameter. Here $\bm{f}^{\leftrightarrow p}$ is the force due to the interaction between particles, and $\bm{f}_p^{\leftarrow f}$ and $\bm{L}_p^{\leftarrow f}$ are the force and momentum due to the fluid-solid interaction, defined as
\begin{subequations}
\begin{align}
 \bm{f}^{\leftarrow f}_p &= \oint_{\partial V_p} \bm{\sigma} \cdot \bm{n} \text{d} A, \\
 \bm{L}^{\leftarrow f}_p &= \oint_{\partial V_p} \bm{r} \times \left( \bm{\sigma} \cdot \bm{n} \right) \text{d}A.
\end{align}
\end{subequations}
Here, $\bm{\sigma} = - p \mathbb{I} + 2 \mu \mathbb{D}$ is the Cauchy stress tensor, where $\mathbb{I}$ is the identity tensor, $\mu$ the kinematic fluid viscosity, and $\mathbb{D}$ the strain rate tensor, and $\bm{n}$ is the unit vector normal to the surface of the particle.

\section{Numerical method}
\label{sec:numerical}

The above-discussed governing equations are numerically integrated using the in-house solver Fujin (\texttt{https://groups.oist.jp/cffu/code}), which is based on an incremental pressure-correction scheme. It considers the Navier--Stokes equations written in primitive variables in a staggered grid, and uses second-order finite differences in all the directions. The Adams-Bashforth time scheme is used for advancing the momentum equation in time. The Poisson equation for the pressure enforcing the incompressibility constraint is solved using a fast and efficient approach based on the Fast Fourier Transform. The solver is parallelised and is based on a Cartesian block decomposition of the domain in the $x$ and $y$ directions, and uses the Message Passing Interface (MPI) library for maximum portability. The governing equations for the particles are dealt with using the immersed boundary method introduced by \citet{hori-rosti-takagi-2022}. The fluid-solid coupling is achieved in an Eulerian framework \citep{kajishima-etal-2001}, and accounts for the inertia of the fictitious fluid inside the solid phase, to properly reproduce their behaviour in both the neutrally-buoyant case and in the presence of density difference between the fluid and solid phases. The soft sphere collision model first proposed by \citet{tsuji-etal-1993} is used to prevent the interpenetration between particles. A fixed-radius near neighbours algorithm \citep[see][and references therein]{monti-etal-2021} is used for the particle interaction to avoid an otherwise prohibitive increase of the computational cost when the number of particles increases. The computational domain consists of a box of size $L$, and is discretised using $N=1024$ equispaced points in the three directions, to ensure that for all cases, all the scales down to the smallest dissipative ones are solved, leading to $\eta/\Delta x= O(1)$, where $\Delta x$ denotes the grid spacing and $\eta$ is the Kolmogorov scale; see the single-phase scale-by-scale budget in the right panel of figure \ref{fig:Spec}. At the initial time, the particles are distributed randomly within the domain.

\subsection{Grid independency}

To demonstrate that the grid resolution is adequate for all particle sizes, an additional simulation has been carried out for the smallest particle size considered $D/L=0.0104$ (or $D/\eta=16$) and $M=0.3$ (see table \ref{tab:simulations}). The number of grid points has been halved, leading to $N=512$ equispaced points in the three directions. This means that the number of grid cells across each particle decreases from $16$ with the standard grid, to $8$ for the coarser grid. 
\begin{figure}
\centering
\includegraphics[width=0.49\textwidth]{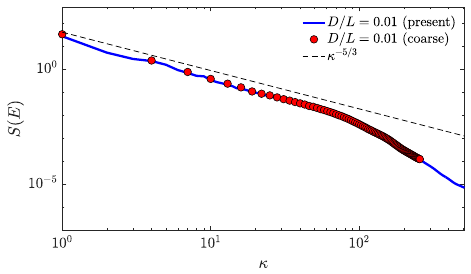}
\caption{Energy spectrum for $D/L = 0.0104$ ($D/\eta=16$) and $M=0.3$. The blue line is for the actual simulation with the $N=1024$ grid. The red circles are for the simulation with the coarser $N=512$ grid.}
\label{fig:SpecD16G5}
\end{figure}
Figure \ref{fig:SpecD16G5} shows that the energy spectra obtained with the two grids agree pretty well at all scales, confirming the adequacy of the considered grid resolution. All the presented results are obtained with the $N=1024$ grid, to properly solve all the scales of the flow down to the Kolmogorov scale (in this case, $\eta/\delta_x = O(1)$). 

\section{List of investigated cases}

Here we provide a list of the configurations that have been considered in the present work. The particle diameter is varied between $0.0207 \le D/L \le 0.0796$, or $16 \le D/\eta \le 123$ (here $\eta$ refers to the single-phase Kolmogorov length-scale). The volume fraction is kept constant to a value of $\Phi_V =V_p/(V_p + V_f) = 0.0792$, where $V_p$ is the total volume of particles, as in \cite{olivieri-cannon-rosti-2022}, being large enough for a non-dilute suspension and small enough for the particle-particle collision to be subdominant. The number of particles $N_p$, thus, is increased as the particle diameter is decreased. The particle density $\rho_p$ is varied between $1.3 \le \rho_p/\rho_f \le 105$ to consider both light and heavy particles, corresponding to a variation of the mass fraction $M = \rho_p V_p/( \rho_p V_p + \rho_f V_f)$ between $ 0.1 \le M \le 0.9$, to a variation of the mass loading $\phi_m = \rho_p V_p/(\rho_f V_f) = M/(1-M)$ between $0.11 \le \phi_m \le 9$, and to a variation of the Stokes number $St$ between $0.2 \lesssim St \lesssim 720$. The Stokes number is defined as $St=\tau_p/\tau_f$ where
\begin{equation*}
\tau_p = \frac{\rho_p}{\rho_f} \frac{D^2}{18 \nu}
\end{equation*}
is the relaxation time of the particle velocity and
\begin{equation*}
\tau_f=\frac{\mathcal{L}}{\sqrt{ 2 \overline{\aver{E}}/3}}
\end{equation*}
is the turnover time of the largest eddies; here
\begin{equation*}
\mathcal{L}=\frac{\pi}{4 \overline{\aver{E}}/3}\int_0^\infty \frac{S(E)}{\kappa} \text{d}\kappa
\end{equation*}
 is the fluid integral scale. Details of the numerical simulations are provided in table \ref{tab:simulations}, together with the corresponding micro-scale Reynolds number $Re_\lambda$ and the Stokes number.

\begin{table}
\caption{Details of the numerical simulations carried out for the present parametric study. $D$ is the particle diameter, $\eta$ the Kolmogorov scale, $N_p$ the number of particles, $\rho_p$ and $\rho_f$ the volumetric density of the fluid and the density of the particles, $M$ the mass fraction, $Re_\lambda$ is the Reynolds number based on $u'=\sqrt{2\overline{\aver{E}}/3}$ and on the Taylor length scale $\lambda$, $St$ is the Stokes number (see text).}
\label{tab:simulations}
\centering
\begin{ruledtabular}
\begin{tabular}{ccccccccccccc}
$D/L$ & & $D/\eta$     & & $N_p$ & & $\rho_p/\rho_f$ & & $M$   & &  $Re_\lambda$ & & $St$ \\
\hline
$-$     & & $ -    $   & & $ - $ & & $ -$            & &  $ -$ & & $435.01$ & & $-$\\
        & &            & &       & &                 & &       & &          & &    \\
$0.0796$   & & $123$   & & $300$ & & $1.3$           & & $0.1$ & & $434.05$ & & $10.39$\\
$0.0796$   & & $123$   & & $300$ & & $4.98$          & & $0.3$ & & $397.47$ & & $41.34$\\
$0.0796$   & & $123$   & & $300$ & & $17.44$         & & $0.6$ & & $347.48$ & & $133.43$\\
$0.0796$   & & $123$   & & $300$ & & $104.69$        & & $0.9$ & & $278.36$ & & $712.69$\\
        & &            & &       & &                 & &       & &          & &\\
$0.0414$    & & $64$   & &$2129$ & & $1.3$           & & $0.1$ & & $428.05$ & & $4.98$   \\
$0.0414$    & & $64$   & &$2129$ & & $4.98$          & & $0.3$ & & $400.94$ & & $9.99$   \\
$0.0414$    & & $64$   & &$2129$ & & $9.518$         & & $0.45$& & $579.15$ & & $21.49$  \\
$0.0414$    & & $64$   & &$2129$ & & $17.44$         & & $0.6$ & & $425.15$ & & $32.19$  \\
$0.0414$    & & $64$   & &$2129$ & & $104.69$        & & $0.9$ & & $225.59$ & & $144.65$ \\
        & &            & &       & &                 & &       & &          & & \\
$0.0207$    & & $32$   & &$17036$& & $1.3$           & & $0.1$ & & $435.36$ & & $0.74$ \\
$0.0207$    & & $32$   & &$17036$& & $4.98$          & & $0.3$ & & $363.32$ & & $2.69$ \\
$0.0207$    & & $32$   & &$17036$& & $9.518$         & & $0.45$& & $564.05$ & & $5.13$ \\
$0.0207$    & & $32$   & &$17036$& & $12.858$        & & $0.525$& & $635.29$& & $7.04$ \\
$0.0207$    & & $32$   & &$17036$& & $17.44$         & & $0.6$ & & $691.96$ & & $9.42$ \\
$0.0207$    & & $32$   & &$17036$& & $34.9$          & & $0.75$& & $382.00$ & & $12.53$ \\
$0.0207$    & & $32$   & &$17036$& & $104.69$        & & $0.9$ & & $186.02$ & & $25.44$ \\
        & &            & &       & &                 & &       & &          & &\\
$0.0104$    & & $16$   & &$136293$& & $1.3$          & & $0.1$ & & $441.61$ & & $0.20$ \\
$0.0104$    & & $16$   & &$136293$& & $4.98$         & & $0.3$ & & $379.86$ & & $0.62$ \\
$0.0104$    & & $16$   & &$136293$& & $17.44$        & & $0.6$ & & $277.74$ & & $1.50$ \\
$0.0104$    & & $16$   & &$136293$& & $104.69$       & & $0.9$ & & $365.34$ & & $8.15$ \\
 \end{tabular}
\end{ruledtabular} 
\end{table}

All simulations have been advanced for approximately $18 L_o/U_o$ time units, after reaching a statistically steady state. For $D/L=0.0207$, or $D/\eta=32$, the simulations have been advanced for a longer period, i.e. between $30 L_o/U_o$ and $60 L_o/U_o$, to ensure convergence of the large-scale statistics (e.g. the temporal mean flow) shown in the discussion.

\section{The direction of the particles' trajectories changes with the initial condition}

\begin{figure}
\centering
\includegraphics[width=0.98\textwidth]{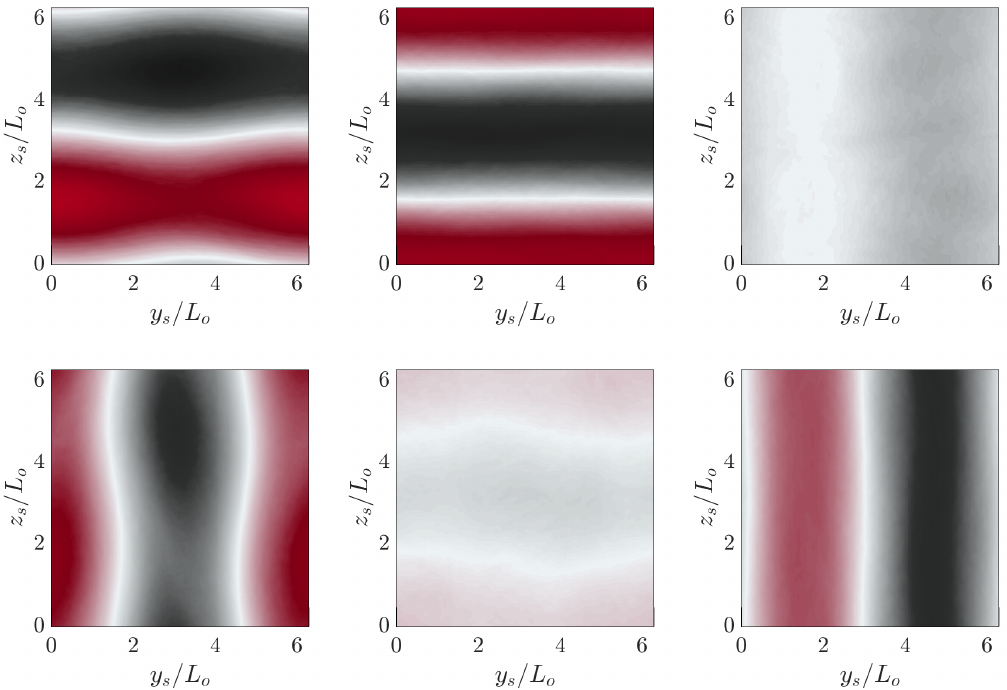}
\caption{Mean flow velocity components for the $D/L=0.0207$ particulate case with $M=0.6$ (top) and $M=0.75$ (bottom) in the $x_s=\pi L_o$ plane, in the Cartesian reference system $(x_s,y_s,z_s)$ of the simulation. From left to right the panels are for $U_s$, $V_s$ and $W_s$, i.e. the components of the mean velocity aligned with three axes. The black-to-red colour map goes from $-9 \le U_i/U_o \le 9$.}
\label{fig:meanflow}
\end{figure}

In this section, we show that the plane where the trajectories of the particles lay in the anisotropic state depends on the initial distribution of the particles. 
 As an example, we consider the $D/L=0.0207$ particulate cases with $M=0.6$ and $M=0.75$. Here we denote with $(x_s,y_s,z_s)$ the Cartesian reference system used in the simulation and with $(x,y,z)$ the Cartesian reference system used for presenting the results; similarly, $(U_s,V_s,W_s)$ and $(U,V,W)$ are the temporal mean velocity components in the two sets of coordinates. Figure~\ref{fig:meanflow} plots the three components of the mean velocity $(U_s,V_s,W_s)$ in the $ x_s = L/2$ plane for $M=0.6$ (top) and $M=0.75$ (bottom). For $M=0.6$, particles are found to move along straight trajectories in the $x_s-y_s$ plane, attenuating thus the $W_s$ component of the mean flow. In contrast, for $M=0.75$, particles are found to move along straight trajectories in the $x_s-z_s$ plane, attenuating the $V_s$ velocity components.

However, for the sake of clarity and uniformity of the results, when presenting the results in the main text, we adopt a Cartesian reference system $(x,y,z)$ with the $z$ direction being orthogonal to the particle trajectories for all cases. In the above $M=0.75$ case, for example, $z_s \rightarrow x$, $x_s \rightarrow y$ and $y_s \rightarrow z$ ($W_s \rightarrow U$, $U_s \rightarrow V$ and $V_s \rightarrow W$).

\section{ABC parameters that approximate the mean flow}

In this section, we show how we evaluate the parameters of the generic ABC flow that approximates the mean flow, $\bm{U}$, of the turbulent single-phase and particulate configurations. The generic $L$-periodic ABC flow can be written as:
\begin{equation}
\bm{u}_{ABC} =V_o
\begin{pmatrix}
A \sin(\frac{2\pi}{L} z) + C \cos( \frac{2 \pi}{L}y) \\
B \sin(\frac{2\pi}{L} x) + A \cos( \frac{2 \pi}{L}z) \\
C \sin(\frac{2\pi}{L} y) + B \cos( \frac{2 \pi}{L}x) \\
\end{pmatrix},
\label{eq:ABCflow}
\end{equation}
where $A$, $B$ and $C$ are bounded between $0$ and $1$ and determine the shape of the flow, and $V_o$ is a constant that determines the flow magnitude. As discussed above, in the particulate cases, we chose the Cartesian reference system such that the $z$ direction is aligned with the velocity component that is mostly attenuated by the particle motion (corresponding to the direction orthogonal to the particle trajectories in the anisotropic state). Without losing generality, thus, we set $A=1$ for all cases and hypothesise that the solid phase modifies the values of $B$, $C$ and $V_o$. The $A$, $B$, $C$ and $V_o$ parameters to use in equation \ref{eq:ABCflow} to approximate $\bm{U}$ are therefore obtained by solving the following system of equations:
\begin{equation}
\begin{cases}
 V_o A  = \frac{1}{L^3} \int_0^{L}  \int_0^{L}  \int_0^{L} \left( U \sin( \frac{2\pi}{L} z) + V \cos( \frac{2\pi}{L}z) \right) \text{d} z  \text{d} y  \text{d} x \\
 V_o B  = \frac{1}{L^3} \int_0^{L}  \int_0^{L}  \int_0^{L} \left( V \sin( \frac{2\pi}{L} x) + W \cos( \frac{2\pi}{L} x) \right) \text{d} z  \text{d} y  \text{d} x \\
 V_o C  = \frac{1}{L^3} \int_0^{L}  \int_0^{L}  \int_0^{L} \left( W \sin( \frac{2\pi}{L}y ) + U \cos( \frac{2\pi}{L}y) \right) \text{d} z  \text{d} y \text{d} x \\
 A = 1
\end{cases}.
\label{eq:ABCeq}
\end{equation}
Note, indeed, that when replacing $(U,V,W)$ with the components of $\bm{u}_{ABC}$, the first three equations are identically satisfied. 

\begin{figure}
\centering
\includegraphics[width=0.8\textwidth]{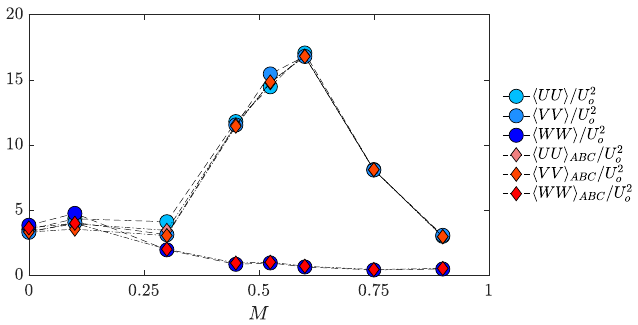}
\caption{Circles: variances of the mean velocity components for the single-phase case and for the $D/L=0.0207$ particulate cases with $ 0.1 \le M \le 0.9$. Diamonds: variances of the components of the generic ABC flow that approximate the single-phase case and the $D/L=0.0207$ particulate cases.}
\label{fig:D32RMS_cmp}
\end{figure}

As an example, figure \ref{fig:D32RMS_cmp} plots the variances of the components of the temporal mean velocity $\bm{U}$ for the $D/L = 0.0207$ particulate cases with $ 0.1 \le M \le 0.9$ and shows that they are in pretty good agreement with the variances of the generic ABC flow obtained with the above-discussed procedure.

\section{ABC Laminar flow and Poincar\'{e} map}

\begin{figure}
\centering
\includegraphics[width=0.8\textwidth]{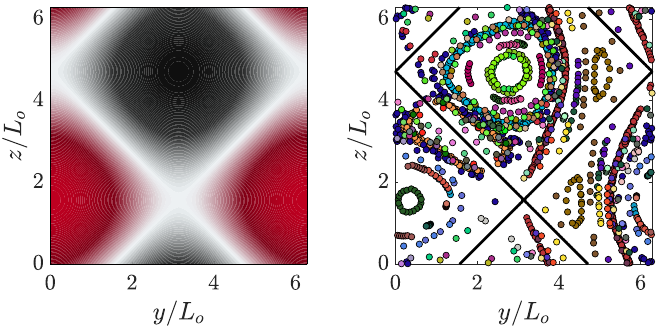}
\caption{Laminar ABC flow $\bm{u}_{ABC}$ with $A=B=C=1$ in the $x/L_o=\pi$ plane. Left: $u_{ABC}$; the symmetric black-to-red comourmap goes from $-2 \le u_{ABC} \le 2$. Right: Poincar\'{e} map; each point of the same colour corresponds to a successive intersection with the $x/L_o = \pi$ plane of the same streamline. The black line is for $u_{ABC}=0$.}
\label{fig:ABCflow}
\end{figure}

Figure \ref{fig:ABCflow} shows the structure of the laminar ABC flow with $A=B=C=1$ in the $x/L_o = \pi$ plane.
The left panel is for $u_{ABC}$. The right panel, instead, shows the $x/L_o=\pi$ Poincar\'{e} section, which describes the Lagrangian structure of the flow. It is a two-dimensional coding of the three-dimensional ABC flow, in which we represent only successive intersections of the flow streamlines with the $x/L_o = \pi$ plane \citep{poincare-1982,dombre-etal-1986}. The density of the points in the Poincar\'{e} map is proportional to the magnitude of the velocity component perpendicular to the plane. Here, indeed, the density of the points is null along the $\sin(z/L_o) + \cos(y/L_o) = 0$ line, where $u_{ABC}=0$.

\section{Flow modulation as a function of $m_p$ and $\phi_m$} 

To provide a more exhaustive picture of the parameters at play, in this section, we show the dependence of the flow modulation on the mass loading and on the mass of the single particle. 

\begin{figure}
\centering
\includegraphics[width=0.98\textwidth]{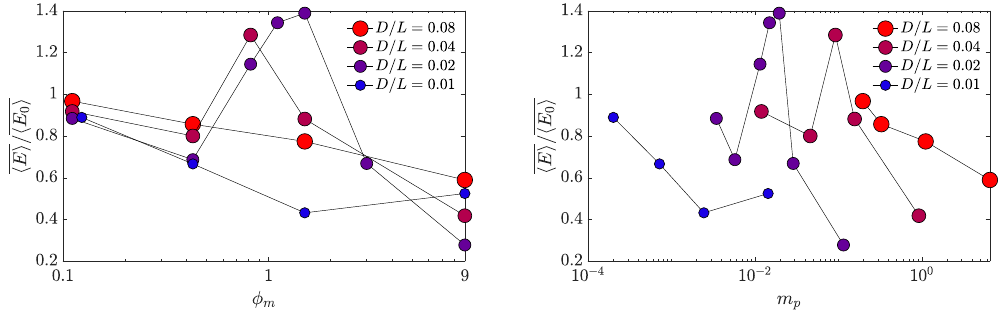}
\caption{Total kinetic energy of the flow as a function of the mass loading (left) and of the dimensionless particle mass $m_p=\rho_p/\rho_f \pi/6 \left( D/L_o \right)^3$ (right).}
\label{fig:Ener_phi_mp}
\end{figure}
Figure \ref{fig:Ener_phi_mp} shows the flow modulation as a function of the mass loading $\phi_m= \rho_p V_p/(\rho_f V_f)$ and of the single-particle mass $m_p=\rho_p/\rho_f \pi/6 \left( D/L_o \right)^3 $. The mass loading is similar to the mass fraction, but is not bounded between $0$ and $1$; the two quantities are related by $\phi_m = M/(1-M)$. The dependence of $\overline{\langle E \rangle}/\overline{\langle E_o \rangle}$ on $\phi_m$ closely resemble the dependence on $M$, with the maximum energy enhancement observed for $\phi_m \approx 1.2$. Regarding $m_p$, instead, the right panel of figure \ref{fig:Ener_phi_mp} shows that the energy enhancement and flow two-dimensionalisation occurs for a specific range of $m_p$, i.e. $10^{-2} \le m_p \le 10^{-1}$. This resembles the dependence of the results on the Stokes number $St$.

\section{On the physical mechanism}

In this section, we shed light on the physical mechanism responsible for the flow two-dimensionalisation. 
\begin{figure}
\centering
\includegraphics[width=0.98\textwidth]{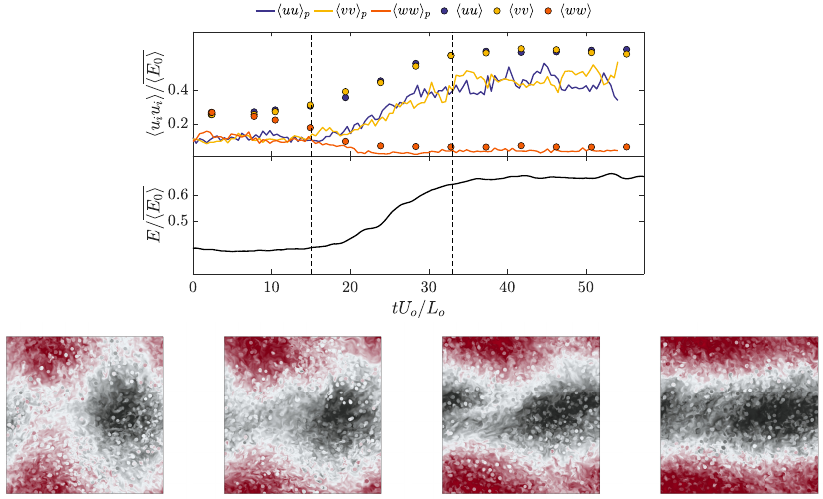}
\caption{Top: Evolution in time of the energy and of the particle and fluid velocity variances for $D/L=0.02$ and $M=0.75$. Bottom: instantaneous fluid velocity snapshots. Time increases from left to right. The left panel is for the isotropic regime at $t \approx 2.5 L_o/U_o$, the right panel is for the anisotropic regime at $t \approx 50 L_o/U_o$, and the central panels are for the transtion between the two regimes at $t \approx 15 L_o/U_o$ and $ t \approx 28 L_o/U_o$.}
\label{fig:Ener_time}
\end{figure}
Figure \ref{fig:Ener_time} shows the time evolution of the anisotropy of the solid and fluid phases for $D/L=0.02$ and $M=0.75$. The flow approaches the isotropic regime before deviating towards the more energetic and anisotropic regime. The top panel details the instantaneous anisotropy of the fluid (symbols) and solid (lines) phases. The time signal shows that the anisotropy arises first in the fluid phase and then in the solid phase: the root mean squares of the three velocity components start differing at $t \approx 10 L_o/U_o$ for the fluid phase, and at $t \approx 15 L_o/U_o$ for the particles. This suggests that the transition from the isotropic state to the anisotropic one is driven by the fluid phase, and not directly by the particle motion. Based on this observation, we speculate that the flow two-dimensionalisation is the result of a bifurcation of the largest scales of the flow, triggered by the perturbations generated by the particles when they are not able to follow the cellular ABC pattern and tend to follow more straight trajectories. For small and large $St$, the flow remains in the isotropic regime, as the perturbation field induced by the solid phase does not have the proper spatial structure to induce this bifurcation. It is worth stressing that, when studying the sequence of bifurcations of the low-$Re$ single-phase ABC flow, \cite{podvigina-pouquet-1994} found in the chaotic regime a similar anisotropic state --- referred to as $\mathcal{A}_2$ --- with $B=C\approx0.36A$, which spontaneously emerges for intermediate $Re$. It is thus possible that the presence of particles and their tendency to follow straight trajectories changes the stability of the system in the $5 \le St \le 60$ and $M > 0.3$ portion of the $M-St$ space of parameters, enabling thus the occurrence of an anisotropic $\mathcal{A}_2$-like state (which indeed favours straight streamlines) also at these larger $Re$.

The role of the inhomogeneous mean shear on the flow two-dimensionalisation has been investigated by considering a different volume forcing to sustain turbulence, that does not generate a coherent motion at the largest scales. In particular, we have carried out additional simulations of the flow in the $L$ triperiodic box using the forcing introduced by \citet{eswaran-pope-1988} to sustain turbulence. For these additional simulations, the case with $D/L=0.0207$ ($D/\eta=32$) is considered and the Reynolds number is slightly smaller: in the single-phase reference case, it is set to $Re_\lambda \approx 330$.
\begin{figure}
\centering
\includegraphics[width=0.49\textwidth]{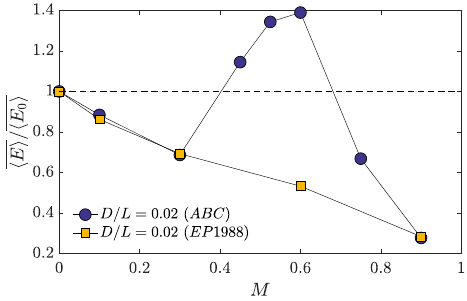}
\caption{Dependence of the flow modulation by particles on the external forcing. The blue circles are for the Arnold-Beltrami-Childress forcing. The yellow squares are for the \cite{eswaran-pope-1988}'s forcing}
\label{fig:forcing}
\end{figure}
Figure \ref{fig:forcing} compares the flow modulation as a function of the mass fraction for the two forcings. The blue circles refer to the results obtained with the ABC forcing, and the yellow squares refer to the random forcing. The two curves collapse pretty well for $M \le 0.3$; here particles modulate all scales in an isotropic way in both cases. For larger $M$ the two curves deviate: in the absence of the inhomogeneous mean shear, the flow energy monotonically decreases with $M$ in agreement with the increase of the inertia of the system, and the energy enhancement associated with the flow two-dimensionalisation is not observed.

\bibliographystyle{apsrev4-1}